\documentclass[twocolumn,showpacs,preprintnumbers,amsmath,amssymb]{revtex4}
\usepackage{graphicx}% Include figure files
\usepackage{dcolumn}% Align table columns on decimal point
\usepackage{bm}% bold math
\usepackage{epsf}

\newcommand{\ul}[1]{\underline{#1}}

\newcommand{\dul}[1]{\underline{\underline{#1}}}

\newcommand{\eps}{\epsilon}

\newcommand{\sinc}{{\rm sinc}}

\newcommand{\bea}{\begin{eqnarray}}

\newcommand{\eea}{\end{eqnarray}}

\begin{document}

\bibstyle{plain}

%\preprint{APS/123-QED}

\title{Spatial Correlation Functions of\\Random Electromagnetic Fields in the Presence of a\\ Semi-Infinite Isotropic Medium
}

\author{Luk R. Arnaut}
 \email{luk.arnaut@npl.co.uk}
\affiliation{%
National Physical Laboratory\\
Division of Enabling Metrology\\
Hampton Road,
Teddington
TW11 0LW\\
United Kingdom
}%

%\date{\nodate}% It is always \today, today,

\begin{abstract}
We extend a previous analysis of spatial correlation functions for classical electromagnetic vector fields near a perfectly conducting boundary [PRE, {\bf 73}, 036604 (2006)] to the case of an isotropic semi-infinite medium with planar interface and characterized by a first-order impedance boundary condition. 
The analytical results are illustrated with calculations for the case of point separations in the direction perpendicular to the interface. 
For the incident plus reflected field, the dependence of the complex-valued and inhomogeneous spatial correlation function on the permittivity, permeability and conductivity of the medium is determined. 
For the refracted field, the spatial correlation is again complex-valued but homogeneous and highly sensitive to the value of the refractive index. 
Based on the derived dependencies, novel nonlocal measurement methods for precision characterization of electromagnetic material properties are suggested.
The influence of the directionality of incidence for electromagnetic beams is investigated. Narrowing the beam width results in a slower decrease of the amplitude of the correlation function as a function of point separation.
Previously obtained asymptotic results for statistically homogeneous random free fields are retrieved as special cases. 
\end{abstract}

\maketitle

\tableofcontents
%\listoffigures

\section{Introduction}
By extending a series of earlier studies for homogeneous free fields \cite{bour1}--\cite{hillv44n2}, we recently derived and analyzed spatial correlation functions of inhomogeneous random classical electromagnetic (EM) fields in \cite{arnaPRE}. 
In that analysis, the configuration consisted of a perfect electrically conducting (PEC) infinite planar boundary, resulting in a superposition of incident and reflected hemispherical statistically isotropic random fields in front of the interface. 
Since a PEC boundary exhibits constant, i.e., angle- and polarization-independent reflection coefficients for both perpendicular (transverse electric, TE) and parallel (transverse magnetic, TM) wave polarizations, the reflection of a statistically isotropic incident field exhibits an isotropic angular spectrum as well. On the other hand, the spatial correlation of the incident plus reflected fields was found to be inhomogeneous (i.e., dependent on the absolute distance of either one of the two point locations with respect to the interface), as a consequence of the statistical field anisotropy imposed by the EM boundary condition at the interface. First-order statistics (probability distributions) of the energy density for this configuration were derived in \cite{arnawall}, which were also found to exhibit inhomogeneity through action at-a-distance.

In this paper, we extend this previous study by considering spatial correlation functions for EM fields in the presence of a semi-infinite isotropic medium, as a second canonical configuration of fundamental interest. 
The impedance boundary condition causes the reflection and transmission coefficients to depend on both the polarization state (TE, TM, or hybrid) and the angle of incidence. As a result, the angular spectra of both the reflected and refracted fields are now no longer hemispherically isotropic, but are nonuniformly weighted across the solid angle of incidence. Secondly, we investigate the influence of directional incidence (sectorial solid angle of incidence centered around a central direction), including narrow EM beams as a limiting case.

The present analysis and results are relevant to several practical problems of interest, e.g., coherence properties of stellar light transmitted through an atmosphere or inside optical instruments, radio waves reflected by the Earth's soil or ionosphere, multipath scattering by man-made objects or precipitation, multi-mode cavities, etc. The field coherency -- which is the basic EM quantity in such scenarios -- is expressed in terms of reflection and transmission coefficients for plane waves impinging onto a single planar interface. By extension, results more general multi-layer configurations are obtained without difficulty, by simply substituting the Fresnel coefficients for a semi-infinite medium with corresponding expressions for stratified media.

\section{Theory}
\subsection{Reflected plus incident fields}
\begin{figure}[htb] \begin{center} 
\begin{tabular}{c}
\ \epsfxsize=5cm \epsfbox{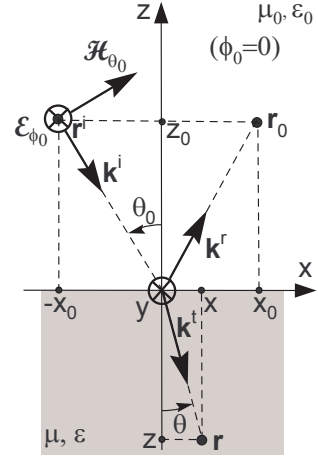}\ \\
\end{tabular}
\end{center}
{ \caption{\label{fig:coordTETM}
Coordinate system and local plane of incidence ($\phi_0=0$, ${\bf 1}_{\phi_0}={\bf 1}_y$) for single TE wave component reflected and refracted by a semi-infinite isotropic medium.
}}
\end{figure}

Consider a semi-infinite isotropic medium with scalar permittivity $\epsilon=\epsilon_r\epsilon_0$, permeability $\mu=\mu_r\mu_0$, conductivity $\sigma$ and first-order surface impedance $\eta=\sqrt{\mu/\epsilon}=\eta_0\sqrt{\mu_r/\epsilon_r}$ where $|\arg(\eta)| \leq \pi/4$.
This medium occupies the half-space $z \leq 0$ (Fig. \ref{fig:coordTETM}). 
We assume a time-harmonic random incident field $({\bf E}^i,{\bf H}^i)$ which can be expanded as an isotropic angular spectrum of plane waves, each specified by a triplet $({\bf \cal E}^i,{\bf \cal H}^i,{\bf k}^i)$ and propagating toward the interface, i.e., ${\bf k}^i\cdot {\bf 1}_z < 0$, 
where $||{\bf k}^i|| \equiv k_0 = \omega\sqrt{\mu_0\eps_0}$ is the (constant) free-space wavenumber of each incident plane wave. A harmonic time dependence $\exp(j\omega t)$ is assumed and suppressed.
The overall incident electric field ${\bf E}^i$ at ${\bf r}={\bf r}^i$ can then be represented as
\bea
{\bf E}^i \left ( {\bf r}^i \right ) = \frac{1}{2\pi} \int \hspace{-2mm} \int_{\Omega_0} {\bf \cal E}^i({ \Omega}) \exp \left ( - j {\bf k}^i \cdot {\bf r}^i \right ) {\rm d}\Omega,
\label{eq:defspectrum}
\eea
The integral (\ref{eq:defspectrum}) is valid for inhomogeneous random fields and hence applicable to the present configuration, unlike homogeneous random fields which, strictly, require a Fourier-Stieltjes representation incorporating generalized fields (distributions).
Incidence and refraction of the plane waves in the upper and lower hemispheres $\Omega_0$ and $\Omega$ is governed by angles $\theta_0$ and $\theta$, respectively, for their propagation direction relative to the surface normal.
The wavenumber within the refracting medium is $ ||{\bf k}^t || \equiv k = \omega \sqrt{\mu\epsilon}=k_0\sqrt{\mu_r \epsilon_r}$. The refracted electric field ${\bf E}({\bf r})$ is expanded in a similar way as (\ref{eq:defspectrum}), mutatis mutandis.

For a general stratified multi-layered medium, including the particular case of a semi-infinite isotropic medium, TE and TM waves constitute a set of uncoupled eigenmodes. Hence, their contributions to the resultant field can be evaluated separately and then superimposed, at any location. We refer to Sections II and III of \cite{arnaPRE} for notations and detailed calculations of the TE/TM decomposition of a random field with respect to the surface normal. 
%****FIGURE 1***
%**************
For an isotropic semi-infinite medium and TE polarization, the resultant (i.e., incident plus reflected) field at ${\bf r}_0(x_0,y_0,z_0)$ within the local plane of incidence $\phi=\phi_0$ is
\bea
{\cal E}_y \exp (-j \bf{k} \cdot \bf{r}_0) 
&=& {\cal E}_{y_0} \exp \left ( - j k_0 \varrho_0 \sin\theta_0 \right ) 
\nonumber\\ &~& \times
\left \{ \left [ 1 + \Gamma_\perp (\theta_0) \right ] \cos \left ( k_0 z_0 \cos \theta_0 \right ) \right.\nonumber\\
&~& \left.
     + j \left [ 1 - \Gamma_\perp (\theta_0) \right ] \sin \left ( k_0 z_0 \cos \theta_0 \right ) 
\right \}
\nonumber\\
\label{eq:EtyTE}
\\
{\cal H}_x \exp (-j \bf{k} \cdot \bf{r}_0) 
&=& \frac{{\cal E}_{y_0}}{\eta_0} \cos \theta_0 \exp \left ( - j k_0 \varrho_0 \sin \theta_0 \right ) 
\nonumber\\ &~& \times
\left \{ \left [ 1 - \Gamma_\perp (\theta_0) \right ] \cos \left ( k_0 z_0 \cos \theta_0 \right )
\right.\nonumber\\
&~& \left.
     + j \left [ 1 + \Gamma_\perp (\theta_0) \right ] \sin \left ( k_0 z_0 \cos \theta_0 \right ) 
\right \}
\nonumber\\
\label{eq:HtxTE}
\\
{\cal H}_z \exp (-j \bf{k} \cdot \bf{r}_0) 
&=& \frac{{\cal E}_{y_0}}{\eta_0} \sin \theta_0 \exp \left ( - j k_0 \varrho_0 \sin \theta_0 \right )
\nonumber\\ &~& \times 
\left \{ \left [ 1 + \Gamma_\perp (\theta_0) \right ] \cos \left ( k_0 z_0 \cos \theta_0 \right )
\right.\nonumber\\
&~& \left.
     + j \left [ 1 - \Gamma_\perp (\theta_0) \right ] \sin \left ( k_0 z_0 \cos \theta_0 \right ) 
\right \}
\label{eq:HtzTE}
\nonumber\\
\eea
where $\varrho_0 = x_0 \cos \phi_0 + y_0 \sin \phi_0$, ${\cal E}_{y_0} = {\cal E}_{\phi_0} \cos \phi_0 + {\cal E}_{\theta_0} \cos \theta_0 \sin \phi_0$, ${\cal E}_{\phi_0} = {\cal E}_0 \cos \psi_0$, ${\cal E}_{\theta_0} = - {\cal E}_0 \sin \psi_0$, in which the random polarization angle $\psi_0$ is uniformly distributed within the local transverse plane spanned by ${\bf 1}_{\phi_0}$ and ${\bf 1}_{\theta_0}$. 
In (\ref{eq:EtyTE})--(\ref{eq:HtzTE}), ${\bf k} = {\bf k}^i$ when ${\bf k} \cdot {\bf 1}_z < 0$ and ${\bf k} = {\bf k}^r = (\dul{\bf I} - 2 {\bf 1}_z {\bf 1}_z) \cdot {\bf k}^i$ for ${\bf k} \cdot {\bf 1}_z > 0$.
The Fresnel TE reflection coefficient is
\bea
\Gamma_\perp(\theta_0) = 
\frac{\eta k u - \eta_0 \sqrt{k^2-k^2_0+k^2_0 u^2}}
     {\eta k u + \eta_0 \sqrt{k^2-k^2_0+k^2_0 u^2}},
\eea
in which $u \stackrel{\Delta}{=} \cos \theta_0$, where $\stackrel{\Delta}{=}$ denotes a definition.
Comparing (\ref{eq:EtyTE})--(\ref{eq:HtzTE}) above with (4)--(5) in \cite{arnaPRE}, it follows that the $z$-dependence of the resultant field is no longer spatially harmonic when $0 \not = \Gamma_{\perp}(\theta_0) \not = \pm 1$, unlike in the case of a PEC surface.
Similarly, for the wave components that are TM with respect to the plane of incidence, we have
\bea
{\cal H}_y \exp (-j \bf{k} \cdot \bf{r}_0) 
&=& - {\cal H}_{y_0} \exp \left ( - j k_0 \varrho_0 \sin \theta_0 \right ) 
\nonumber\\ &~& \times
\left \{ \left [ 1 - \Gamma_\parallel (\theta_0) \right ] \cos \left ( k_0 z_0 \cos \theta_0 \right )\right.\nonumber\\
&~& \left.
     + j \left [ 1 + \Gamma_\parallel (\theta_0) \right ] \sin \left ( k_0 z_0 \cos \theta_0 \right ) 
\right \}
\nonumber\\
\label{eq:HtyTM}
\\
{\cal E}_x \exp (-j \bf{k} \cdot \bf{r}_0) &=& 
\eta_0 {\cal H}_{y_0} \cos \theta_0 \exp \left ( - j k_0 \varrho_0 \sin\theta_0 \right ) 
\nonumber\\ &~& \times
\left \{ \left [ 1 + \Gamma_\parallel (\theta_0) \right ] \cos \left ( k_0 z_0 \cos \theta_0 \right )\right.\nonumber\\
&~& \left.
     + j \left [ 1 - \Gamma_\parallel (\theta_0) \right ] \sin \left ( k_0 z_0 \cos \theta_0 \right ) 
\right \}
\nonumber\\
\label{eq:EtxTM}
\\
{\cal E}_z \exp (-j \bf{k} \cdot \bf{r}_0) &=& 
\eta_0 {\cal H}_{y_0} \sin \theta_0 \exp \left ( - j k_0 \varrho_0 \sin\theta_0 \right ) 
\nonumber\\ &~& \times
\left \{ \left [ 1 - \Gamma_\parallel (\theta_0) \right ] \cos \left ( k_0 z_0 \cos \theta_0 \right )\right.\nonumber\\
&~& \left.
     + j \left [ 1 + \Gamma_\parallel (\theta_0) \right ] \sin \left ( k_0 z_0 \cos \theta_0 \right ) 
\right \}
\label{eq:EtzTM}
\nonumber\\
\eea
with ${\cal H}_{y_0} = {\cal H}_{\phi_0} \cos \phi_0 + {\cal H}_{\theta_0} \cos \theta_0 \sin \phi_0$, ${\cal H}_{\phi_0} = {\cal H}_0 \sin \psi_0$, ${\cal H}_{\theta_0} = {\cal H}_0 \cos \psi_0$, ${\cal H}_0 = {\cal E}_0 / \eta_0$, and Fresnel TM reflection coefficient 
\bea
\Gamma_\parallel(\theta_0) =
\frac{\eta \sqrt{k^2-k^2_0+k^2_0 u^2} - \eta_0 k u}
     {\eta \sqrt{k^2-k^2_0+k^2_0 u^2} + \eta_0 k u}
\eea
Whilst we shall limit the further analysis to point separations in normal direction ($\Delta r {\bf 1}_r = \Delta z {\bf 1}_z$), the results are easily extended to arbitrary directions using the methodology outlined in Section IV of \cite{arnaPRE}.
 
For the TE waves, substitution of (\ref{eq:EtyTE}) into (\ref{eq:defspectrum}), evaluated at two locations ${\bf r}_{1,2}=z_{1,2}{\bf 1}_z$ for ${\bf r}_{0}$, enables the calculation of $E_y ({\bf r}_1) \cdot E^*_y({\bf r}_2)$ via double integration with respect to corresponding ranges $\Omega_1$ and $\Omega_2$ \cite[eq. (16)]{arnaPRE}, where the asterisk denotes complex conjugation. This is followed by ensemble averaging of this product, assuming delta-correlated random field components \cite[eq. (17)]{arnaPRE}, i.e.,  
$ \langle {\bf \cal E}_1 (\Omega_1) \cdot {\bf \cal E}^*_2 (\Omega_2) \rangle 
\equiv 
\langle {\cal E}_{1\theta} (\Omega_1) \cdot {\bf \cal E}^*_{2\theta} (\Omega_2) \rangle 
+ 
\langle {\cal E}_{1\phi}   (\Omega_1) \cdot {\bf \cal E}^*_{2\phi}   (\Omega_2) \rangle 
= 
2 C \delta [(\Omega_1 \cup \Omega_2) \setminus (\Omega_1 \cap \Omega_2)]$, where $C \stackrel{\Delta}{=} \langle |{\cal E}_0|^2 \rangle/4$.
If, in addition, each complex Cartesian component exhibits a zero mean value, then these impositions on the first- and second-order moments define, unambiguously, a 3-D complex (6-D real) multivariate Gauss normal distribution with independent and identically distributed components ${\bf \cal E}_{i\alpha}$. The results also apply to more general distributions for ${\bf \cal E}_{1,2}$, provided that their first- and second-order moments satisfy the stated expressions.
It follows that for a general isotropic impedance boundary condition, the TE field coherency $\langle E_y(z_1) E^*_y(z_2) \rangle$ can be written as a sum of four terms, viz., 
\bea
\langle E_y (z_1) E^*_y(z_2) \rangle = I_{y1} + I_{y2} + I_{y3} + I_{y4}
\label{eq:Iy}
\eea
where
\bea
I_{y1} &=& 2C
\int^1_0
\left | 1 - \Gamma_\perp (u ) \right |^2 
\nonumber\\
&~&\times
\sin \left ( k_0 z_1 u \right ) \sin \left ( k_0 z_2 u \right )
{\rm d}u\label{eq:Iy1TE}\\
I_{y2} &=& 2C
\int^1_0
\left | 1 + \Gamma_\perp (u ) \right |^2 
\nonumber\\
&~&\times
\cos \left ( k_0 z_1 u \right ) \cos \left ( k_0 z_2 u \right )
{\rm d}u\\
I_{y3} &=& j 2C
\int^1_0
\left [ 1 - \Gamma_\perp (u ) \right ] \left [ 1 + \Gamma^*_\perp (u ) \right ] 
\nonumber\\
&~&\times
\sin \left ( k_0 z_1 u \right ) \cos \left ( k_0 z_2 u \right )
{\rm d}u
\\
I_{y4} &=& - j 2C
\int^1_0
\left [ 1 + \Gamma_\perp (u) \right ] \left [ 1 - \Gamma^*_\perp (u) \right ] 
\nonumber\\
&~&\times
\cos \left ( k_0 z_1 u \right ) \sin \left ( k_0 z_2 u \right )
{\rm d}u\label{eq:Iy4TE}
\eea
Note that $I_{y3}\not = I^*_{y4} $ and $\Im(I_{y3}+I_{y4}) \not = 0$ unless $z_1 = z_2$, so that the spatial coherencies are in general complex-valued.
The integrals (\ref{eq:Iy1TE})--(\ref{eq:Iy4TE}) evaluate to closed-form but cumbersome expressions.
Expressions for the TM coherencies follow in an analogous manner by substituting (\ref{eq:EtxTM}) and (\ref{eq:EtzTM}) into (\ref{eq:defspectrum}), yielding
\bea
\langle E_x (z_1) E^*_x(z_2) \rangle = I_{x1} + I_{x2} + I_{x3} + I_{x4}
\label{eq:Ix}
\eea
where
\bea
I_{x1} &=& 2C
\int^1_0
\left | 1 - \Gamma_\parallel (u ) \right |^2 
\nonumber\\
&~&\times
u^2
\sin \left ( k_0 z_1 u \right ) \sin \left ( k_0 z_2 u \right )
{\rm d}u\label{eq:Ix1TM}\\
I_{x2} &=& 2C
\int^1_0
\left | 1 + \Gamma_\parallel (u ) \right |^2 
\nonumber\\
&~&\times
u^2
\cos \left ( k_0 z_1 u \right ) \cos \left ( k_0 z_2 u \right )
{\rm d}u\\
I_{x3} &=& j 2C
\int^1_0
\left [ 1 - \Gamma_\parallel (u ) \right ] \left [ 1 + \Gamma^*_\parallel (u ) \right ] 
\nonumber\\
&~&\times
u^2
\sin \left ( k_0 z_1 u \right ) \cos \left ( k_0 z_2 u \right )
{\rm d}u
\\
I_{x4} &=& - j 2C
\int^1_0
\left [ 1 + \Gamma_\parallel (u) \right ] \left [ 1 - \Gamma^*_\parallel (u) \right ] 
\nonumber\\
&~&\times
u^2 
\cos \left ( k_0 z_1 u \right ) \sin \left ( k_0 z_2 u \right )
{\rm d}u\label{eq:Ix4TM}
\eea
and
\bea
\langle E_z (z_1) E^*_z(z_2) \rangle = I_{z1} + I_{z2} + I_{z3} + I_{z4}
\label{eq:Iz}
\eea
where
\bea
I_{z1} &=& 2C
\int^1_0
\left | 1 + \Gamma_\parallel (u ) \right |^2 
\nonumber\\
&~&\times \left ( 1 - u^2 \right )
\sin \left ( k_0 z_1 u \right ) \sin \left ( k_0 z_2 u \right )
{\rm d}u\label{eq:Iz1TM}\\
I_{z2} &=& 2C
\int^1_0
\left | 1 - \Gamma_\parallel (u ) \right |^2 
\nonumber\\
&~&\times \left ( 1 - u^2 \right )
\cos \left ( k_0 z_1 u \right ) \cos \left ( k_0 z_2 u \right )
{\rm d}u\\
I_{z3} &=& j 2C
\int^1_0
\left [ 1 + \Gamma_\parallel (u ) \right ] \left [ 1 - \Gamma^*_\parallel (u ) \right ] 
\nonumber\\
&~&\times \left ( 1 - u^2 \right )
\sin \left ( k_0 z_1 u \right ) \cos \left ( k_0 z_2 u \right )
{\rm d}u
\\
I_{z4} &=& - j 2C
\int^1_0
\left [ 1 - \Gamma_\parallel (u) \right ] \left [ 1 + \Gamma^*_\parallel (u) \right ] 
\nonumber\\
&~&\times \left ( 1 - u^2 \right ) 
\cos \left ( k_0 z_1 u \right ) \sin \left ( k_0 z_2 u \right )
{\rm d}u\label{eq:Iz4TM}
\eea
For the normal field ${\bf E}_z = E_z {\bf 1}_z$, the tangential field ${\bf E}_t = E_x {\bf 1}_x + E_y {\bf 1}_y$, and the total, i.e., vector field ${\bf E} = E_x {\bf 1}_x + E_{y} {\bf 1}_y + E_z {\bf 1}_z$, we have
\bea
\langle E_z (z_1) E^*_z (z_2) \rangle &=& 
\sum^4_{\ell=1} I_{z\ell},\label{eq:EzIalpha}\\
\langle E_t (z_1) E^*_t (z_2) \rangle &=& 
\sum_{\alpha=x,y} \sum^4_{\ell=1} I_{\alpha\ell},\label{eq:EtIalpha}\\
\langle E (z_1) E^* (z_2) \rangle &=& 
\sum_{\alpha=x,y,z} \sum^4_{\ell=1} I_{\alpha\ell}\label{eq:EIalpha}
\eea

\subsection{Refracted fields}
For the field transmitted (refracted) across the interface, we obtain in an analogous manner, with the aid of the field transmission coefficients $T_{\perp,\parallel}(\theta_0)$ and Snell's law $k_0 \sin \theta_0 = k \sin \theta$, 
\bea
&~&\langle E_y (z_1) E^*_y(z_2) \rangle =
2C \int^1_0 |T_\perp(u)|^2 
\nonumber\\
&~&~~\times
\exp \left [ j k_0 (z_1-z_2) \sqrt{ \left ( \frac{k}{k_0} \right )^2 - 1 + u^2} \right ] {\rm d}u
\\
%\eea
%\bea
&~&
\langle E_x (z_1) E^*_x(z_2) \rangle = 
2C \int^1_0 |T_\parallel(u)|^2 
\nonumber\\
&~&~~\times
\left [ 
1 - 
\left ( \frac{k_0}{k} \right )^2 + 
\left ( \frac{k_0}{k} \right )^2 u^2 
\right ]
\nonumber\\
&~&~~\times
\exp \left [ j k_0 (z_1-z_2) \sqrt{ \left ( \frac{k}{k_0} \right )^2 - 1 + u^2} \right ] {\rm d}u
\\
%\eea
%\bea
&~& \langle E_z (z_1) E^*_z(z_2) \rangle = 
2C \int^1_0 |T_\parallel(u)|^2 \left ( \frac{k_0}{k} \right )^2 
\left ( 1 - u^2 \right )
\nonumber\\
&~&~~\times
\exp \left [ j k_0 (z_1-z_2) \sqrt{ \left ( \frac{k}{k_0} \right )^2 - 1 + u^2} \right ] {\rm d}u
\eea
where the Fresnel TE and TM transmission coefficients are
\bea
T_\perp(u) =  1+\Gamma_{\perp}(u) = \frac{2 \eta k u}{\eta k u + \eta_0 \sqrt{k^2 - k^2_0 + k^2_0 u^2} },
\eea
\bea
T_{\parallel}(u) &=& \frac{\cos \theta_0}{\cos \theta} \left [ 1+\Gamma_{\parallel}(u) \right ] \nonumber\\
&=& \frac{2 \eta k u}{\eta_0 k u + \eta \sqrt{k^2 - k^2_0 + k^2_0 u^2} }
\eea
respectively. The coherencies now exhibit complex-harmonic dependencies on the separation distance only, as in the case of free random fields in an infinite homogeneous medium. Thus, unlike for the incident plus reflected field, the spatial correlation of the refracted field is {\em homogeneous}, i.e., dependent on $k\Delta z = k|z_1-z_2|$ only. Physically, this is a consequence of the fact that no interference exists beyond the interface. Nevertheless, because of the $\theta_0$-dependence of $T_{\perp,\parallel}$, the coherency of the refracted field is different from that of the incident field.

\section{Numerical results}
\subsection{Good conductor}
For good but imperfect nonmagnetic conductors ($\sigma \gg \omega \epsilon_0$, $\epsilon = \epsilon_0$, $\mu=\mu_0$), we can approximate $\eta/\eta_0 \simeq \sqrt{\omega \eps_0/(2\sigma)} (1\pm j)$ and $\theta \simeq 0$. The reflection coefficients then  become
\bea
\Gamma_\perp (u)     \simeq \frac{\eta u - \eta_0            }{\eta u + \eta_0  },~~~
\Gamma_\parallel (u) \simeq \frac{\eta             - \eta_0 u}{\eta   + \eta_0 u}
\eea
Upon substituting these expressions into (\ref{eq:Iy1TE})--(\ref{eq:Iy4TE}), (\ref{eq:Ix1TM})--(\ref{eq:Ix4TM}) and (\ref{eq:Iz1TM})--(\ref{eq:Iz4TM}), followed by a transition to the limit $\sigma / (\omega \epsilon_0) \rightarrow +\infty$, it is verified that for a PEC surface only the terms $I_{\alpha 1}$ in (\ref{eq:EzIalpha})--(\ref{eq:EIalpha}) are nonzero, for either polarization.

\begin{figure}[htb] \begin{center} \begin{tabular}{c}
\ \epsfxsize=8cm\epsfbox{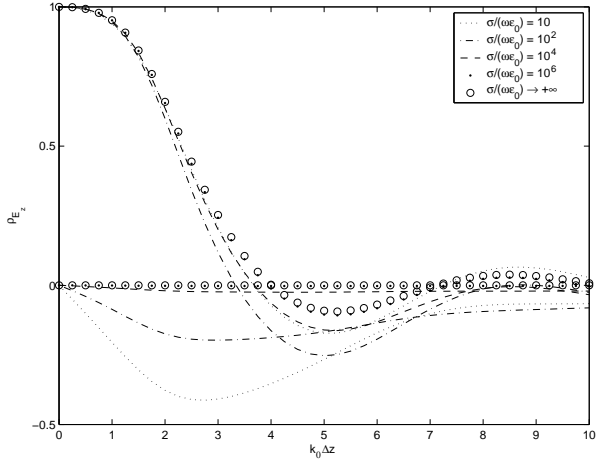}\ \\
\end{tabular}
\end{center}
{\bf \caption{\label{fig:spatcorrnearconduc_param_sigma_kz0_0p7854_Enorm}
Scf of the incident plus reflected normal field $E_z$ for selected values of $\sigma/(\omega\eps_0)$ at $k_0 z_0=\pi/4$ as a function of separation $k_0  \Delta z$ in normal direction.
Curves originating at ordinate value $1$ represent $\Re[\rho_{E_z}(k_0 \Delta z)]$; curves originating at ordinate value $0$ represent $\Im[\rho_{E_z}(k_0 \Delta z)]$.
}}
\end{figure}

\begin{figure}[htb] \begin{center} \begin{tabular}{c}
\ \epsfxsize=8cm\epsfbox{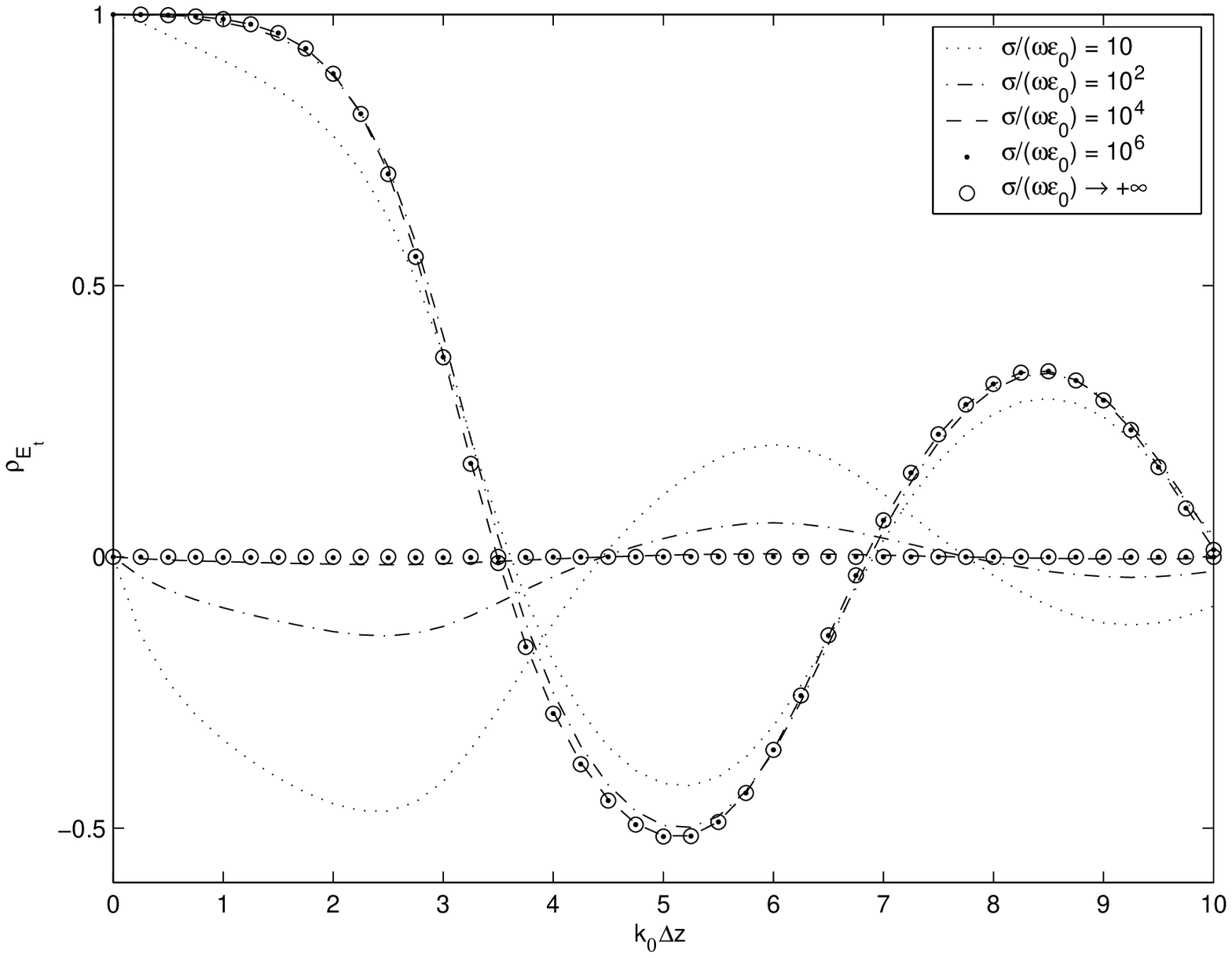}\ \\
\end{tabular}
\end{center}
{\bf \caption{\label{fig:spatcorrnearconduc_param_sigma_kz0_0p7854_Etan}
Scf of the incident plus reflected tangential field $E_t$ for selected values of $\sigma/(\omega\eps_0)$ at $k_0 z_0=\pi/4$ as a function of separation $k_0  \Delta z$ in normal direction.
Curves originating at ordinate value $1$ represent $\Re[\rho_{E_t}(k_0 \Delta z)]$; curves originating at ordinate value $0$ represent $\Im[\rho_{E_t}(k_0 \Delta z)]$.
}}
\end{figure}

\begin{figure}[htb] \begin{center} \begin{tabular}{c}
\ \epsfxsize=8cm\epsfbox{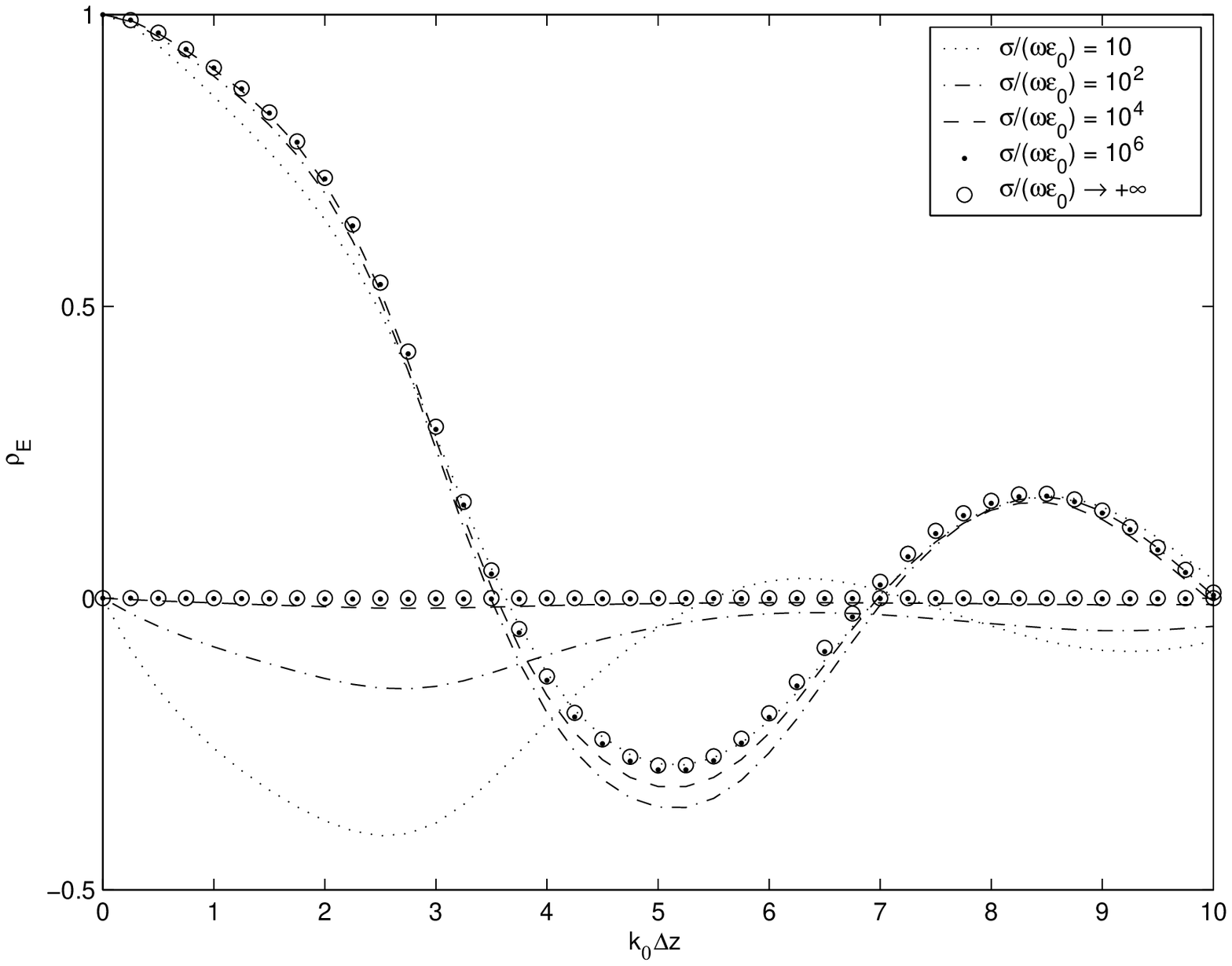}\ \\
\end{tabular}
\end{center}
{\bf \caption{\label{fig:spatcorrnearconduc_param_sigma_kz0_0p7854_Etot}
Scf of the incident plus reflected total (vector) field $E$ for selected values of $\sigma/(\omega\eps_0)$ at $k_0 z_0=\pi/4$ as a function of separation $k_0  \Delta z$ in normal direction.
Curves originating at ordinate value $1$ represent $\Re[\rho_{E}(k_0 \Delta z)]$; curves originating at ordinate value $0$ represent $\Im[\rho_{E}(k_0 \Delta z)]$.
}}
\end{figure}

Figure \ref{fig:spatcorrnearconduc_param_sigma_kz0_0p7854_Enorm} compares the spatial correlation function (scf) of the normal component of the incident plus reflected electric field
\bea
&~& \rho_{E_z} (k_0 \Delta z; k_0 z_0) \stackrel{\Delta}{=} 
\nonumber\\
&~&~~~
\frac{\langle E_z(k_0 z_0) E^*_z(k_0 z_0+k_0 \Delta z) \rangle
     }
     { \sqrt{ \langle |E_z(k_0 z_0)|^2 \rangle \langle |E_z(k_0 z_0+k_0 \Delta z)|^2 \rangle }
     }
\eea
at $k_0 z_0=\pi/4$ for selected values of $\sigma/(\omega\eps_0) \gg 1$ with the corresponding function for a PEC surface. Finite values of $\sigma/(\omega\eps_0)$ are seen to cause the first zero crossing of $\Re[\rho_{E_z}(k_0 \Delta z)]$ to occur at smaller values of $k_0 \Delta z$ compared to a PEC boundary. Scfs for $E_t$ and $E$ (Figs. \ref{fig:spatcorrnearconduc_param_sigma_kz0_0p7854_Etan} and \ref{fig:spatcorrnearconduc_param_sigma_kz0_0p7854_Etot}) show that corresponding differences for $\Re[ \rho_{E_{(t)}}(k_0 \Delta z) ]$ between finitely conducting and PEC surfaces are less pronounced than for $E_z$. 
Compared to a PEC surface, the damping of $\Re[\rho_{E_t}(k_0 \Delta z)]$ is qualitatively different from that for $\Re[\rho_{E_{z}}(k_0 \Delta z)]$. 
Also, finite conductivities yield nonvanishing imaginary parts of the scf, indicating that $E(k_0 z_0)$ and $E^*(k_0 z_0+k_0 \Delta z)$ or their components are, on average, no longer in phase. 
This effect can be exploited as a means to measure surface conductivity. 
%***************
%****FIGURE 2***
%***************
%***************
%****FIGURE 3***
%***************
%***************
%****FIGURE 4***
%***************

\clearpage

\subsection{Lossless isotropic dielectric medium}
As a second special case, we analyze the effect of the permittivity of a lossless isotropic dielectric medium on the scf. For brevity, we now limit the presentation to results for the amplitude of the vector field $E$ only.

\begin{figure}[htb] \begin{center} \begin{tabular}{c}
\ \epsfxsize=8cm\epsfbox{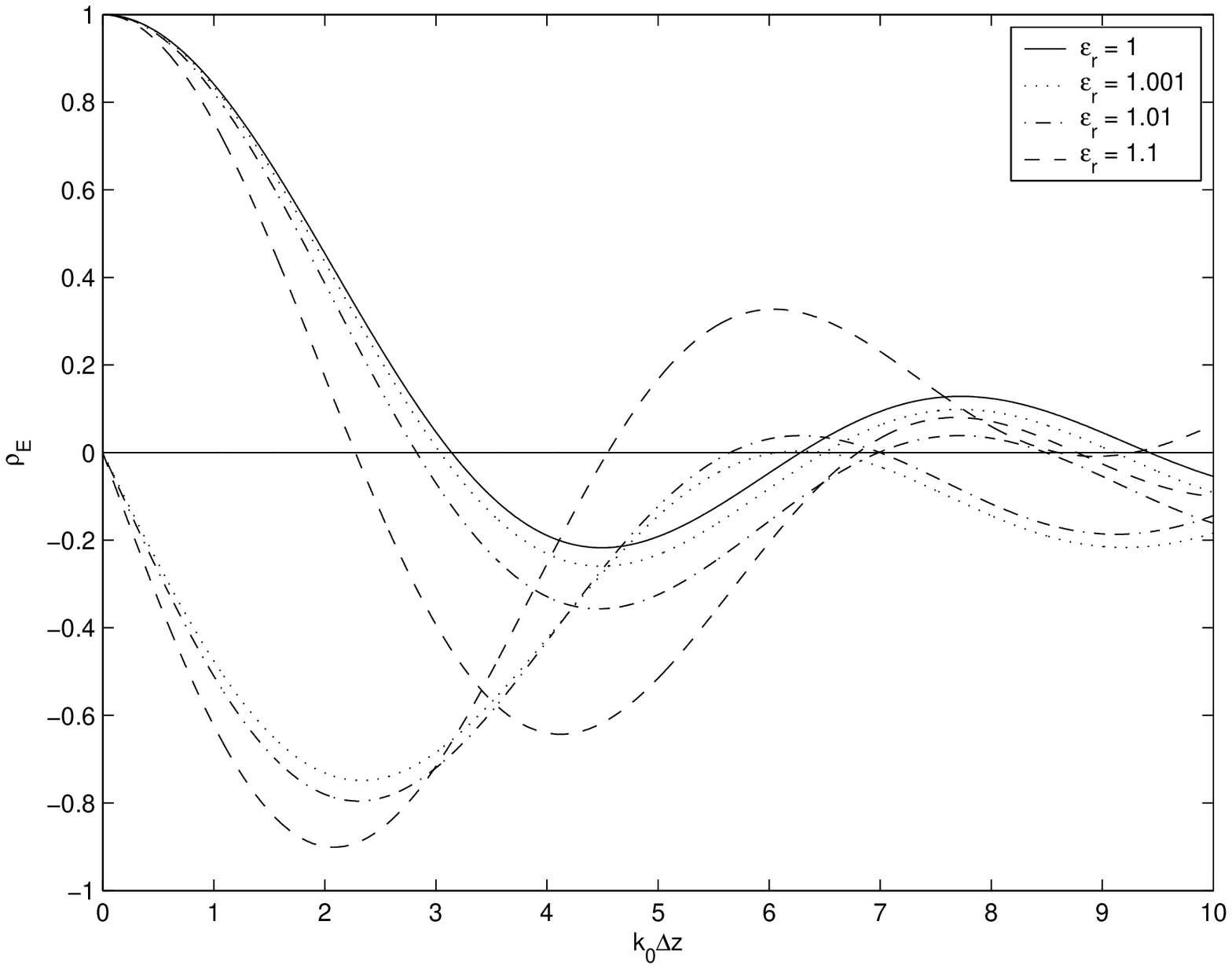}\ \\
\end{tabular}
\end{center}
{\bf \caption{\label{fig:spatcorrneardielec_Tx_param_epsr_Etot}
Scf of the refracted vector field $E$ for selected values of $\epsilon_r=\eps/\eps_0$ at arbitrary $k_0 z_0$ as a function of separation $k_0 \Delta z$ in normal direction. 
Curves originating at ordinate value $1$ represent $\Re[\rho_E(k_0 \Delta z)]$; curves originating at ordinate value $0$ represent $\Im[\rho_E(k_0 \Delta z)]$.  
}}
\end{figure}

For the refracted field,
Fig. 
\ref{fig:spatcorrneardielec_Tx_param_epsr_Etot}
shows that the permittivity manifests itself by a decrease of the first zero crossing distance (correlation length)  for $\Re[\rho_{E}(k_0 \Delta z)]$, with associated shifts of the local maximum and minimum values toward lower values of $k_0 \Delta z$. Also, the amplitudes of $\Re[\rho_E(k_0 \Delta z)]$ and $\Im[\rho_E(k_0 \Delta z)]$ increase with increasing $\epsilon/\epsilon_0$.
Similar findings apply to $\rho_{E_z}(k_0 \Delta z)$ and $\rho_{E_t}(k_0 \Delta z)$ (not shown).
The high sensitivity of $\rho_E(k_0\Delta z)$ to the value of $\epsilon/\epsilon_0$, in combination with its insensitivity to $k_0z_0$, suggest that measurements of $\rho_E(k_0\Delta z)$ may be used as a precision method for determining the refractive index of a transparent substance.
%***************
%****FIGURE 5***
%***************

\begin{figure}[htb] \begin{center} \begin{tabular}{c}
\ \epsfxsize=8cm\epsfbox{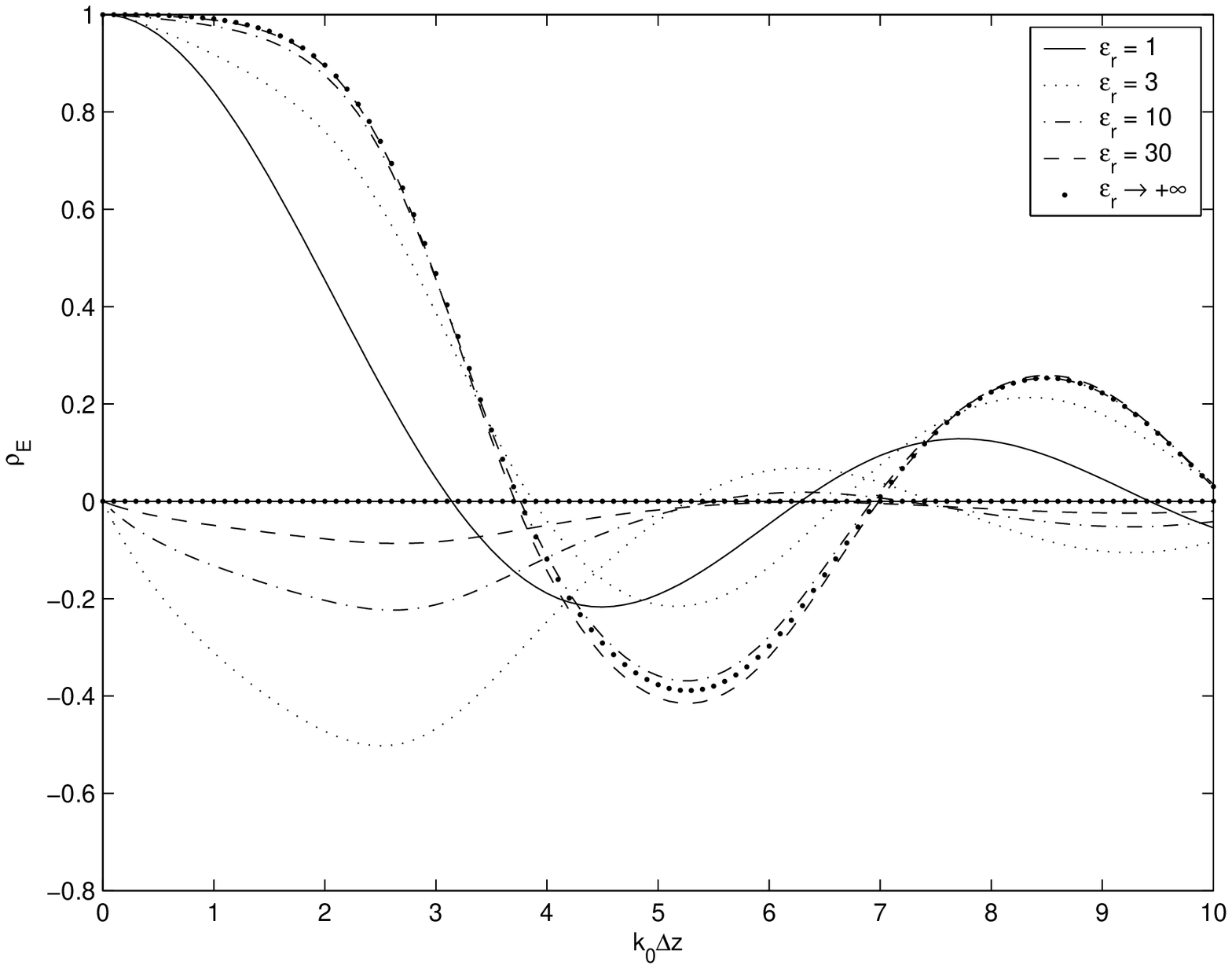}\ \\
\end{tabular}
\end{center}
{\bf \caption{\label{fig:spatcorrdielec_Rx_param_epsr_kz0_0p7854_Etot}
Scf of incident plus reflected vector field $E$ for selected values of $\eps_r=\eps/\eps_0$ at $k_0 z_0 = \pi/4$ as a function of separation $k_0 \Delta z$. Curves originating at ordinate value $1$ represent $\Re[\rho_{E}(k_0 \Delta z)]$; curves originating at ordinate value $0$ represent $\Im[\rho_{E}(k_0 \Delta z)]$.  
}}
\end{figure}

For the incident plus reflected field, Fig. \ref{fig:spatcorrdielec_Rx_param_epsr_kz0_0p7854_Etot} shows $\rho_{E}(k_0 \Delta z;k_0 z_0)$ in the half-space of incidence at $k_0 z_0=\pi/4$, for selected values of $\eps_r \equiv \epsilon/\epsilon_0 $. Comparing the asymptotic curve for $\eps_r \rightarrow +\infty$ with Fig. \ref{fig:spatcorrnearconduc_param_sigma_kz0_0p7854_Etot} for a PEC surface, it is noticed that $\rho_E(k_0 \Delta z)$ is qualitatively similar, but quantitative differences exist, particularly for $k_0 \Delta z \leq 1$. Thus, the scf for the reflected field can be used to distinguish between conducting and high-k dielectric media, both of which exhibit high reflectivities making them otherwise difficult to discern in scalar local measurements.
%***************
%****FIGURE 6***
%***************

\subsection{Directional incidence}
So far, the direction of incidence $(\theta_0, \phi_0)$ of the random field onto the interface was assumed to be uniform within the upper hemisphere ($\Omega_0 = 2 \pi$ sr, viz., $-\pi/2 < \theta_0 < \pi/2$, $0 \leq \phi_0 < \pi$).
In practice, particularly in millimeter-wave and optical regimes, the wavevectors of the incident EM beams are often confined to be within a narrower solid angle 
$\theta_0 - \Delta \theta_0 \leq \theta_0 < \theta_0 + \Delta \theta_0$, 
$\phi_0 - \Delta \phi_0     \leq \phi_0   < \phi_0 + \Delta \phi_0$ and scaling by $2\Delta \phi_0/\pi$ in azimuthal direction.
For $\Delta \theta_0, \Delta \phi_0 \rightarrow 0$, this approaches an unpolarized EM beam incident along the ($\theta_0,\phi_0$)-direction. 

The spatial coherence along the reflected or refracted beam can be calculated as before, upon replacing the integrations $\int^1_0 {\rm d}u$ for the angular spectral averages by $\int^1_{\cos(\Delta \theta_0)}{\rm d}u$.
Along the direction of specular reflection, $-\theta_0$, the tangential and normal point separations are related by $\Delta x /\Delta z = \tan \theta_0 \equiv \sqrt{(1/u)^2 -1}$; along the direction of refraction, $\theta$, their ratio is 
$\Delta x / \Delta z = -\tan \theta \equiv - \sqrt{1-u^2}/\sqrt{(k/k_0)^2 - 1 + u^2}$.
In general, narrow incident and reflected beams fields do not interfere unless incidence is sufficiently close (with respect to the beam width) to the surface normal, whence the solid angles of the incident and reflected waves overlap partially or completely. 

\begin{figure}[htb] \begin{center} \begin{tabular}{c}
\ \epsfxsize=8cm\epsfbox{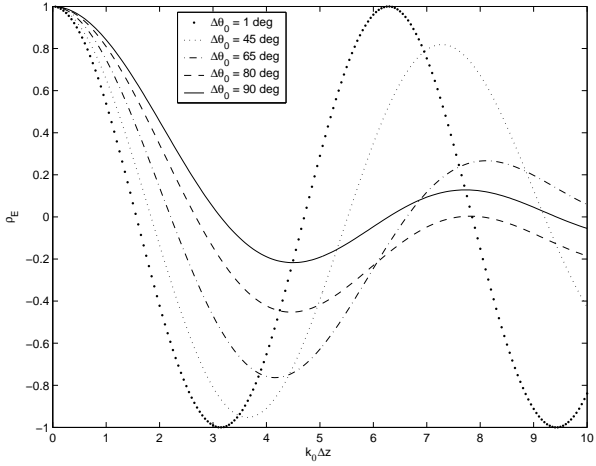}\ \\
\end{tabular}
\end{center}
{\bf \caption{\label{fig:spatcorrnearPEC_param_alpha_kz0_infty_Etot}
Scf of incident plus reflected vector field $E$ in normal direction in front of a PEC surface, at selected values of $\Delta \theta_0$ for $k_0 z_0 \rightarrow +\infty$ as a function of separation $k_0 \Delta z$.
}}
\end{figure}

To illustrate the effect of directionality of incidence on the scf, we consider an incident random field represented by an angular spectrum of elevational width (field of view) $2\Delta \theta_0$ centered around the normal direction ($\theta_0=0$) with preservation ofthe  azimuthal symmetry around this direction ($\Delta \phi_0 = \pi$) and random polarization ($0 \leq \psi_0 < 2\pi$). This corresponds to incidence from within a solid angle $2\pi\Delta \theta_0$ sr.
For a PEC surface, Fig. \ref{fig:spatcorrnearPEC_param_alpha_kz0_infty_Etot} shows $\rho_{E}(k_0 \Delta z)$ at selected values of $\Delta \theta_0$ for $k_0 z_0\rightarrow +\infty$. It can be verified that for $k_0 z_0\rightarrow +\infty$,
\bea
\rho_{E}(k_0 \Delta z; \Delta \theta_0 \rightarrow 0) &\rightarrow &
\rho_{E_t}(k_0 \Delta z; \Delta \theta_0 \rightarrow 0) = 
\nonumber\\
&~&~~~~
\cos(k_0 \Delta z)
\eea 
On the other hand, for $k_0 z_0\rightarrow 0$, 
\bea
\rho_{E}(k_0 \Delta z; \Delta \theta_0 \rightarrow 0) &\rightarrow& 
\rho_{E_t}(k_0 \Delta z; \Delta \theta_0 \rightarrow 0) = \nonumber\\
&~&~~~~
{\rm sgn}[\cos(k_0 \Delta z)]
\label{eq:blockwave}
\eea

%***************
%****FIGURE 7***
%***************

\begin{figure}[htb] \begin{center} \begin{tabular}{c}
\ \epsfxsize=8cm\epsfbox{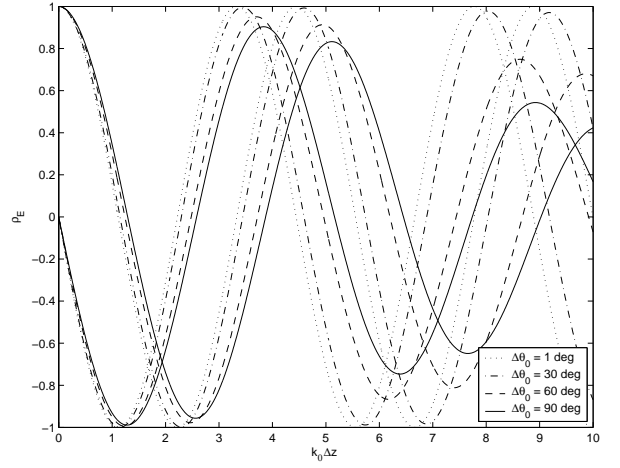}\ \\
\end{tabular}
\end{center}
{\bf \caption{\label{fig:spatcorrneardielec_Tx_param_alpha_epsr_2}
Scf of refracted vector field $E$ at selected values of $\theta_0$ with $\epsilon/\epsilon_0 = 2$ as a function of separation $k_0 \Delta z$ in normal direction.
Curves originating at ordinate value $1$ represent $\Re[\rho_E(k_0 \Delta z)]$; curves originating at ordinate value $0$ represent $\Im[\rho_E(k_0 \Delta z)]$.  
}}
\end{figure}

Fig. \ref{fig:spatcorrneardielec_Tx_param_alpha_epsr_2} shows corresponding results for refraction by an isotropic dielectric medium with $\epsilon/\epsilon_0 = 2$, demonstrating qualitatively similar features for both $\Re[\rho_{E}(k_0 \Delta z)]$ and $\Im[\rho_{E}(k_0 \Delta z)]$, with $\rho_{E}(k_0 \Delta z) \rightarrow \exp(-j k_0 \sqrt{\epsilon_r}\Delta z)$ for $\Delta \theta_0 \rightarrow 0$. In general, a medium with larger $\epsilon_r$ yields more rapidly modulated oscillations of its scf.

%***************
%****FIGURE 8***
%***************

\begin{figure}[htb] \begin{center} \begin{tabular}{c}
\ \epsfxsize=8cm\epsfbox{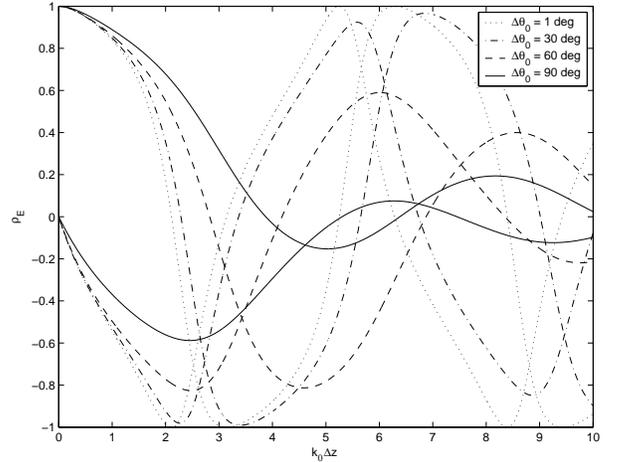}\ \\
\end{tabular}
\end{center}
{\bf \caption{\label{fig:spatcorrneardielec_Rx_param_alpha_epsr_2_kz0_0p7854}
Scf of incident plus reflected vector field $E$ at selected values of $\Delta \theta_0$ with $k_0 z_0=\pi/4$ and $\epsilon/\epsilon_0 = 2$ as a function of separation $k_0 \Delta z$ in normal direction.
Curves originating at ordinate value $1$ represent $\Re[\rho_E(k_0 \Delta z)]$; curves originating at ordinate value $0$ represent $\Im[\rho_E(k_0 \Delta z)]$.  
}}
\end{figure}

Corresponding functions for the incident plus reflected field at $k_0 z_0=\pi/4$ are shown in Fig. \ref{fig:spatcorrneardielec_Rx_param_alpha_epsr_2_kz0_0p7854}.
A general feature is that the oscillations of the scf become less regular when $k_0 z_0$ decreases for a given value $\Delta \theta_0$, or vice versa; see, for example, the plot of $\rho_{E}(k_0 \Delta z)$ for $\Delta \theta_0 = 1$ deg in Fig. \ref{fig:spatcorrneardielec_Rx_param_alpha_epsr_2_kz0_0p7854}. In the limit $k_0 z_0 \rightarrow 0$, the scf tends again to the complex-harmonic square-wave function ${\rm sgn}[\exp(-jk_0 \Delta z)]$.

%***************
%****FIGURE 9***
%***************

\clearpage

\section{Conclusion}
The reflection and transmission properties of an isotropic magneto-dielectric medium have been shown to have a significant effect on the spatial correlation functions of normal, tangential and total EM random vector fields, as a consequence of the EM boundary conditions. An analysis of the corresponding effects on the probability distribution of the energy density of the total field for this configuration will be presented in a forthcoming paper.

The effect of changes in the constitutive parameters on the correlation length is in general ambiguous, because a decreasing first-zero crossing distance of the scf is usually accompanied by an increase in its amplitude, $|\rho_E(k_0 \Delta z)|$, so that e.g. the two definitions investigated in Section V of \cite{arnaPRE} yield diverging tendencies for such changes.

Since the above formulation is in terms of TE and TM reflection and transmission coefficients for plane waves, the analysis can be extended without effort to investigate the scf for more general stratified configurations with uncoupled eigenpolarizations of this kind, e.g., multi-layered stratified media as well as uniaxial anisotropic media, by substituting the appropriate functional forms of these coefficients.

\section{Acknowledgement}
This work was supported by the 2003--2006 Electrical Programme of the National Measurement System Policy Unit of the U.K. Department of Trade and Industry.

\end{document}